\documentclass[11pt,a4paper]{article}
\pdfoutput=1 
\usepackage[utf8]{inputenc}
\usepackage{authblk}

\usepackage{libertine}

\usepackage[margin=1in]{geometry}
\usepackage{amsfonts}
\usepackage{amsmath}
\usepackage{amssymb}
\usepackage{amsthm}
\usepackage{float}
\usepackage[font=small,labelfont=bf]{caption} 
\allowdisplaybreaks
\usepackage{bbold}
\usepackage{hyperref}

\def\fp/{\textup{\textsf{FP}}}
\def\p/{\textup{\textsf{P}}}
\def\np/{\textup{\textsf{NP}}}
\def\conp/{\textup{\textsf{co-NP}}}
\def\fnp/{\textup{\textsf{FNP}}}
\def\tfnp/{\textup{\textsf{TFNP}}}
\def\ptfnp/{\textup{\textsf{PTFNP}}}
\def\ppa/{\textup{\textsf{PPA}}}
\def\ppad/{\textup{\textsf{PPAD}}}
\def\ppads/{\textup{\textsf{PPADS}}}
\def\ppp/{\textup{\textsf{PPP}}}
\def\pwpp/{\textup{\textsf{PWPP}}}
\def\pls/{\textup{\textsf{PLS}}}
\def\cls/{\textup{\textsf{CLS}}}
\def\ppadpls/{\textup{$\textsf{PPAD} \cap \textsf{PLS}$}}
\def\ppapls/{\textup{$\textsf{PPA} \cap \textsf{PLS}$}}
\def\eopl/{\textup{\textsf{EOPL}}}
\def\sopl/{\textup{\textsf{SOPL}}}
\def\ueopl/{\textup{\textsf{UEOPL}}}
\def\fixp/{\textup{\textsf{FIXP}}}
\def\bu/{\textup{\textsf{BU}}}
\def\bbu/{\textup{\textsf{BBU}}}
\def\linearfixp/{\textup{\textsf{Linear-FIXP}}}
\def\pspace/{\textup{\textsf{PSPACE}}}

\begin{document}

\title{\bf Multi-Agent Systems for \\ Computational Economics and Finance}

\author{Michael Kampouridis\thanks{Email: \href{mailto:mkampo@essex.ac.uk}{\texttt{\nolinkurl{mkampo@essex.ac.uk}}}} 
\qquad
Panagiotis Kanellopoulos\thanks{Email: \href{mailto:panagiotis.kanellopoulos@essex.ac.uk}{\texttt{\nolinkurl{panagiotis.kanellopoulos@essex.ac.uk}}}}
\qquad
Maria Kyropoulou\thanks{Email: \href{mailto:maria.kyropoulou@essex.ac.uk}{\texttt{\nolinkurl{maria.kyropoulou@essex.ac.uk}}}} \\\vspace{4pt}
Themistoklis Melissourgos\thanks{Email: \href{mailto:themistoklis.melissourgos@essex.ac.uk}{\texttt{\nolinkurl{themistoklis.melissourgos@essex.ac.uk}}}}
\qquad
Alexandros A. Voudouris\thanks{Email: \href{mailto:alexandros.voudouris@essex.ac.uk}{\texttt{\nolinkurl{alexandros.voudouris@essex.ac.uk}}}}}

\affil{}

\date{\small School of Computer Science and Electronic Engineering \\ University of Essex}

\maketitle

\begin{abstract}
In this article we survey the main research topics of our group at the University of Essex. Our research interests lie at the intersection of theoretical computer science, artificial intelligence, and economic theory. In particular, we focus on the design and analysis of mechanisms for systems involving multiple strategic agents, both from a theoretical and an applied perspective.
We present an overview of our group's activities, as well as its members, and then discuss in detail past, present, and future work in multi-agent systems.
\end{abstract}

\section{The Group} \label{sec:intro}
The School of Computer Science and Electronic Engineering at the University of Essex (UoE) is the home of many researchers actively working on the broad field of Artificial Intelligence. We are members of the Centre for Computational Finance and Economic Agents (CCFEA) in UoE, and our research spans various core areas of multi-agent systems (MAS), focusing primarily at the intersection of theoretical computer science, artificial intelligence, finance and economics. Specifically, our expertise is on algorithmic game theory, computational social choice, machine learning and financial engineering. 
We have a strong publication track record that demonstrates our close working relationship, as well as our national and international collaborations. Our relevant research is published in highly respected international peer-reviewed journals and conferences, such as AAAI, IJCAI, AAMAS, EC, ICALP, STOC, FOCS, Artificial Intelligence Journal, Journal of Artificial Intelligence Research, Games and Economic Behavior, and more. We also participate as Program Committee members in renowned international conferences and have participated in the organisation of several international conferences and workshops, including the 15th International Symposium on Algorithmic Game Theory (SAGT 2022) and the 2022 IEEE Symposium on Computational Intelligence for Financial Engineering and Economics (CIFEr). 

We remark that our work is intertwined with industrial applications and we have been involved in four Knowledge Transfer Partnerships funded by Innovate UK (total value exceeding £1M), as well as an ESRC-funded project with Essex County Council. For example, in collaboration with a large healthcare provider, we are studying more efficient ways to elicit preferences, match healthcare workers and clients, and increase the number of shifts that are covered. Our approach relies heavily on game theory, mechanism design, and machine learning. In another case involving a leading law firm, we worked towards developing a pricing tool that predicts the cost of a requested piece of legal work and allocates the most suitable lawyers to it, based on historical data. This is a fascinating and innovative research direction linking once again machine learning and algorithmic game theory.

\section{Core Areas of our Research} \label{sec:core}
The study of multi-agent systems is a core area of Artificial Intelligence and many different perspectives and methodologies have been adopted, each contributing to the development of MAS. One key area our group has focused on is the {\em existence, computation and quality of equilibria} in environments where multiple intelligent agents act selfishly aiming to optimize their personal agenda, which may collide with the interests of others or the system designer's objectives. Another class of problems we have extensively studied falls within the area of {\em computational social choice}, where agents express preferences over different outcomes and our goal is to make social decisions based on these preferences that satisfy desired properties, such as {\em utilitarian efficiency} or {\em fairness criteria}. We have studied game-theoretic and social choice settings in which the main objective is to provide the agents with the necessary incentives to avoid strategic manipulations, an area known as {\em mechanism design}. Finally, we have considered \emph{financial} applications both through the aforementioned frameworks as well as by using Machine Learning. 
In the rest of this section we provide background on these topics, discuss our contribution, and identify interesting directions for future research.

\subsection{Existence, Computation, and Quality of Equilibria} \label{sec:games}

{\em Strategic games} emerge naturally in environments where rational and intelligent individuals participate. Games model situations where selfish entities aim to maximize their own payoff without caring about that of their competitors.
Informally, a game consists of a set of {\em agents}, each having a set of available \emph{actions}. In such an MAS, a combination of the agents' actions yields an \emph{outcome} which is experienced by all agents; each agent has her own \emph{preference} over the outcomes, often expressed through a valuation or utility function. 
A probability distribution over an agent's actions is called a \emph{strategy}, while a combination of agent strategies in which no agent would get a better (in expectation) outcome by unilaterally deviating to a different strategy is called a \emph{Nash equilibrium} \cite{Nash51}. Such notions of stability often serve as predictions on the outcomes of the corresponding game, thus, have received a great deal of attention from the community. Various other equilibrium notions have also been considered as solution concepts, depending on the game at hand. 

There are three main questions regarding equilibria in games: \emph{Is an equilibrium guaranteed to exist?}, \emph{how fast can we compute one (if it exists)?}, and \emph{how good for the society are the equilibria reached compared to a centrally enforced optimal state?} The last question gives rise to the notions of \emph{price of anarchy} and \emph{price of stability} that quantify how bad the worst (best, respectively) Nash equilibrium is when compared to the state maximizing the aggregate welfare of the society (optimal \emph{social welfare}). We have studied these questions in many different and challenging types of games. 
In what follows, we present some of our relevant work grouped by similarity of model. 

\paragraph{Auctions.}
In a typical auction environment, there is a set of items for sale
and a set of potential buyers. The possible actions of the buyers consist of the potential bids they can place in the auction, and the game is defined by a specific resolution method of the bids, which, given the bids, determines the outcome.
Consider, for example, the generalized second price (GSP) auction, which is the primary auction used for monetizing the use of the Internet.   
In a quite influential paper (see \cite{CKKK+15}) we studied the space of equilibria under GSP and quantified the efficiency loss that can arise under a wide range of sources of uncertainty. 
We also focused on the revenue of the GSP auction, and studied a Bayesian setting wherein the valuations of advertisers are drawn independently from a common regular probability distribution \cite{CKKK14}.

In a combinatorial auction, an outcome consists of an
allocation of the items to the agents and a price for each agent's bundle. 
One of the most prominent equilibrium notions for auctions, where we require per-item prices, is that of a 
\emph{Walrasian equilibrium}, which occurs when no agent prefers any other bundle for the given set of prices and the market clears. Walrasian equilibria have the additional desirable property of maximizing the social welfare, however they are not known to exist even for very restricted valuations, as we showed in \cite{DMS21}.
In the same paper, we presented 
some special cases for which existence of an equilibrium is guaranteed and provided polynomial time algorithms to compute it.

Furthermore, we have analyzed the efficiency of auctions in settings where the bidders have {\em budget constraints} that limit the amount of payments they can afford. More specifically, we have considered auctions for the allocation of divisible resources with bidders that have concave valuation functions for fractions of the resources~\cite{CV2021budgets}, position auctions (including the GSP auction mentioned above)~\cite{voudouris2019position}, and simultaneous combinatorial auctions with payment rules that are convex combinations of the bids~\cite{voudouris2020combinatorial}. 

We have also studied a series of questions related to the existence, efficiency and computational complexity of equilibria in two-stage games modelling settings where multiple vendors compete with each other by offering similar products to unit-demand buyers who in turn have different preferences over the vendors~\cite{caragiannis2017pricing}.

\paragraph{Congestion games.}
In a \emph{congestion game} there is a set of resources and each agent can use a subset of them, possibly under some restrictions. Each resource incurs a cost that is a function of the number of agents that use it. Congestion games have been studied extensively due to the fact that they capture many real-life applications and since they admit {\em pure Nash equilibria}. 

In \cite{cfkk+11} we completely characterized the impact of selfishness and greediness in singleton congestion games, where each strategy consists of a single resource. In particular, we presented tight bounds on the price of anarchy, as well as on the competitiveness of the greedy algorithm for the online setting, when the objective is to minimize the total cost of all agents, as well as a tight upper bound on the price of stability of congestion games with linear cost functions. We have also introduced a setting where agents are partially altruistic and partially selfish and we have determined the impact of this behavior on the overall system performance \cite{CKKK+10}. In an attempt to mitigate the impact of selfish behavior, we studied the question of how much the performance can be improved if agents are forced to pay taxes for using resources, for the setting of atomic congestion games with linear cost functions \cite{CKKK10,CKK08}.

A well-known special case of congestion games is that of task scheduling. 
In \cite{GK19} we considered the problem of scheduling tasks on strategic unrelated machines when no payments are allowed, under the objective of minimizing the makespan. We adopted the monitoring paradigm where a machine is bound by her declarations and provided a (non-truthful) randomized algorithm whose pure price of anarchy is considerably better than the approximation ratio that can be achieved by any truthful mechanism. 

Scheduling tasks to machines is also the focus of \cite{CKKP08}. Each task has a particular latency requirement at each machine and may choose either to be assigned to some machine in order to get serviced provided that her latency requirement is met, or not to participate in the assignment at all. From a global perspective, 
one would aim to maximize the number of tasks that participate in the assignment. However, tasks may behave selfishly aiming to  
get serviced by some machine where her latency requirement is met with no regard to overall system performance. We modelled this selfish behavior as a strategic game, showed how to compute pure Nash equilibria efficiently, and assessed the impact of selfishness on system performance. 

\paragraph{Games on graphs.} 
{\em Segregation} has been the subject of multidisciplinary research for more than half a century, with most work focusing on Thomas Schelling's model~\cite{Schelling69,Schelling71} involving 
two different types of agents. Each agent occupies a node on a graph and is happy if a least some fraction of her neighbors are of the same type; an unhappy agent either jumps to an unoccupied node or swaps locations with another unhappy agent. 
In recent work, we considered Schelling's model through the lens of game theory, by assuming that agents
choose their next location to maximize the fraction of same-type neighbors. We have studied questions related to the existence, computation, and quality of equilibria for many variants of Schelling games 
~\cite{elkind2019jump,kanellopoulos2021modified,kanellopoulos2021strangers,
agarwal2020swap}. Besides this, we have also studied the complexity of computing welfare-maximizing allocations
~\cite{bullinger2021welfare}.

Regarding the classical \emph{stable matching} problem, we considered its generalization that allows for cardinal preferences (as opposed to ordinal) and fractional matchings (as opposed to integral) \cite{caragiannis2021fractional}. In this cardinal setting, stable fractional matchings can have much larger social welfare than stable integral ones. Our goal was to understand the computational complexity of finding an optimal (i.e., welfare-maximizing) stable fractional matching. We considered both exact and approximate stability notions, and provided simple approximation algorithms with weak welfare guarantees; somewhat surprisingly, achieving better approximations is computationally hard.

{\em Opinion formation} is another important topic, where the objective is to study how the opinion of an individual is affected by 
her social acquaintances. In this context, we have studied the existence and quality of equilibria in games consisting of agents, each of whom has a personal belief that can be represented by a real number, but may express a different opinion aiming to minimize the maximum distance of her opinion from her belief and the most distant opinion expressed by any of her friends \cite{caragiannis2022opinion}.

In graph-based \emph{hedonic games}, a group of utility-maximizing agents have hedonic preferences over the agents' set, and wish to be partitioned into clusters so that they are grouped together with agents they prefer. The agents correspond to nodes in a connected graph and their preferences are defined so that shorter graph distance implies higher preference. In \cite{KKPP21} we focused on fractional hedonic games and social distance games and we proved improved bounds on the price of stability.  

\paragraph{Security games.}
To decide the optimal strategy to commit to, the leader in a Stackelberg game can use a variety of learning algorithms that operate by querying the best responses or the payoffs of the follower. Consequently, the follower can potentially deceive the leader by responding as if her payoffs are much different than what they actually are. In \cite{birmpas2021deception}, we proved that it is always possible for the follower to compute payoffs, and thus deceive the leader, that yield near-optimal utility, for various scenarios about the learning interaction between leader and follower. In \cite{elkind2021recounting}, we considered a two-stage district-based voting manipulation scenario which leads to a Stackelberg game. First, an attacker attempts to manipulate the election outcome in favor of a preferred candidate by changing the vote counts in some districts. Afterwards, a defender has the ability to demand a recount in a subset of the manipulated districts, thus restoring the vote counts therein to the original, true values. 

An attacker-defender setting was also studied in \cite{ADMS21} where multiple attackers try to inflict damage to nodes on a network, and a single defender tries to protect the nodes by probabilistically patrolling induced connected subgraphs of the network. In that work we showed a general LP-based algorithm for computing Nash equilibria and then focused on special cases of graphs. First we characterized equilibria, and analyzed some of their properties, like the
\emph{defense ratio}; that is, the ratio of all attackers over the expected number of the ones caught by the defender in equilibrium. We also computed the (parameterized) value of the \emph{price of defense}, i.e., the worst defense ratio over all possible networks.

Investigating a significantly challenging aspect of cryptocurrency, in \cite{KKKT16} we focused on the strategic considerations of miners participating in bitcoin's protocol. We formulated and studied the stochastic game underlying such strategic considerations and showed that, when each miner's computational power is relatively small, their best response matches the expected behavior of the bitcoin designer. However, when a miner's computational power is large, she deviates from the expected behavior, and other Nash equilibria arise. This work has served as a starting point for a long line of works on blockchain protocols.  

\paragraph{Existence and tractability of equilibria.}
Proving existence of Nash equilibria in games with infinite number of strategies is another interesting avenue which requires techniques tailored to the problem at hand. In \cite{CMS18, CMS18-SAGT} we studied such a game in the context of contention resolution over multiple-access transmission channels in communication networks. We extended the single channel setting, and showed existence of anonymous equilibrium protocols which we explicitly described for various feedback models.

As a side quest of their main subject, almost all of the works in this section have studied the question of tractability of equilibria. In other words, they have categorized the equilibrium computation problems according to how fast algorithms they (can) admit under the Turing machine model. \cite{MS17} solely studied this question on evolutionarily stable strategies, and showed that deciding existence of such strategies is \conp/-hard for a wide range of games in nature. Furthermore, in \cite{DFHM2022-pure-circ} we studied various \ppad/-complete problems in algorithmic game theory, and showed that they attain their computational hardness even for very large constant approximation.

On the bright side, we have managed to show (quasi) polynomial time approximation schemes for a significantly broad family of problems in algorithmic game theory and beyond. In \cite{DFMS22}, we presented a master algorithm which finds approximate solutions to non-linear programs with non-linear constraints. Our algorithm is simple and, in particular, of the brute force kind. Given that the domain is succinctly representable -- as is the case for the strategy set in a great number of games -- the algorithm is efficient. Importantly, it is able to handle classical optimization problems such as multi-objective non-linear optimization, and for some families of these problems performs at least as well as the currently best algorithms. Nonetheless, it can be considered a unification of older optimization methods under a single approximation scheme with remarkable simplicity. Finally, a notable implication of our work is that if our algorithm terminates without an approximate solution, then the problem does not admit even an exact solution.

\paragraph{Future directions.} 
Even though there have been many results on the computational complexity of the aforementioned problems by now, there is still a vast, unexplored area for research. The focus of the computer science community has been almost entirely on the worst-case complexity and only little is known about the average case or similar. This raises the following very important open question: \emph{What is the average or smoothed complexity of algorithms for these problems?} Furthermore, as blockchain technology conquers more and more fields, and more protocols are being designed based on, it is important to analyze the theoretical properties of such protocols and the incentives of the miners therein. Finally, a recent series of papers have studied questions about the manipulability of machine learning algorithms for regression or classification, and have shown bounds on the price of anarchy and stability of the induced games. The limits of this research direction have not yet been reached and there are many interesting settings to be considered.

\subsection{Computational Social Choice} \label{sec:comsoc}
Another area of MAS research that we have studied in recent years is that of {\em computational social choice}~\cite{comsoc-book}. One of the most prominent models within this area is that of {\em utilitarian voting}, where we have a set of {\em agents} and a set of {\em alternatives}. Every agent has a private valuation function, which maps the alternatives to numerical values, indicating how much the agent likes each alternative. Due to cognitive limitations, it is typical for each agent to express their preferences using only {\em limited} information, typically by ranking the alternatives from the one she considers the best to the one she considers the worst. A {\em mechanism} (or {\em voting rule}) takes as input the information provided by the agents about their preferences and outputs a subset of alternatives. The goal is to design mechanisms that make decisions to (approximately) maximize some objective function of the values of the agents for the selected alternatives, such as the {\em social welfare} (the total value of the agents) or the {\em egalitarian welfare} (the minimum value over all agents). As mechanisms only have limited information about the valuation functions of the agents the choices they can make are inevitably suboptimal. This loss of efficiency is measured by the {\em distortion}, that is, the worst-case ratio between the maximum possible objective value over all feasible subsets of alternatives and the objective value of the subset chosen by the mechanism. 

Our work in utilitarian voting has focused both on improving the state of the art (see our recent survey for an overview \cite{distortion-survey}), as well as introducing new agendas and opening avenues for future research. In single-winner voting (where the goal is to choose a single alternative), we initiated the study of {\em query-enhanced mechanisms} that take as input the rankings over alternatives provided by the agents, and also have the capability of making a small number of {\em queries} to (approximately) learn parts of the valuations, which they can then use to make better-informed decisions. For this class of mechanisms, we showed tradeoffs between the necessary amount of information (number of queries) to achieve low distortion~\cite{amanatidis2021peeking}. In follow-up work \cite{amanatidis2022matching,amanatidis2022dice}, we showed how our methodology can be adapted and extended to capture several other fundamental social choice models beyond utilitarian voting, including the well-known {\em one-sided matching} problem (in which there is a set of agents that have preferences over a same-sided set of items and the objective is to match each agent to a single distinct item so as to maximize the total value of the agents) and several generalizations of it, such as {\em constrained resource allocation} (where each agent can be matched to more than one item), {\em two-sided matching} (in which each agent has preferences over all other agents and the goal is to create pairs of agents with maximum value for one another), and {\em graph matching} (where the preferences of the agents for other agents are partial and defined according to a directed graph). 

In another line of work, we initiated the study of {\em distributed} voting settings~\cite{filos2020distributed,FV2021approximate,anshelevich2022distributed}. In contrast to centralized voting where there is a single pool of agents and we obtain direct access to the information provided by them, in distributed voting the agents are partitioned into {\em disjoint groups} and the mechanisms work in two steps: We first decide a {\em representative alternative} for each group by holding a local election for the agents therein, and then choose one of these representative alternatives as the overall winner. This model captures important applications such as district-based presidential elections, and many other scenarios in which the agents might be grouped together because of logistic or geographic reasons. 
Recently, we also initiated the study of multiwinner voting (where the goal is to choose some fixed-size subset of alternatives) in settings where the agents have metric preferences over the alternatives (that is, agents and alternatives are represented as points in some --- possibly multidimensional --- metric space, i.e., the distances between agents and alternatives that define the valuations satisfy the triangle inequality)~\cite{caragiannis2022multiwinner}.

\paragraph{Future directions.} 
There are plenty of interesting directions for future work on utilitarian voting. One possible direction would be to combine the models discussed above and investigate the interplay between information and distortion in several single-winner and multiwinner distributed voting, as well as other important settings beyond these, such as {\em participatory budgeting} (where each alternative is associated with a cost and the goal is to choose a subset of alternatives of total cost within a given budget). While the distortion of ordinal voting rules is well-understood in most important settings by now, the known impossibilities do not rule out that machine learning methods could help in improving the efficiency in an instance-by-instance case. In fact, recent work has shown that it is possible to teach machine learning methods to approximately mimic some voting rules such as scoring rules~\cite{anil2021learning}. While this by itself is not groundbreaking, it indicates that it might be possible to learn voting rules that cannot be described through a simple formula or well-defined instructions. So, machine learning has the potential to allow us to learn voting rules that approximately maximize social welfare.

\subsection{Fair Division} \label{sec:fair}
An important class of MAS problems in computer science is that of {\em fair division}, or in other words, fair resource allocation. While such problems may also fall within the general umbrella of computational social choice, the amount of attention they have received in the last century, and particularly in the last couple of decades, justifies discussing them in isolation. In a fair division problem, there is a set of agents that typically have cardinal preferences over a set of items, where these items can correspond to divisible or indivisible goods or chores. The goal is to compute an allocation of the items to the agents such that every agent considers the allocation to be fair. What fairness means can have multiple interpretations, which is why many different fairness notions have been proposed and studied in the literature. The most compelling and widely adopted ones are those of {\em envy-freeness}, which demands that no agent envies what another agent has been allocated, and {\em proportionality}, which demands that every agent obtains at least her proportional share of all items.

Motivated by work on price of anarchy and stability (see Section~\ref{sec:games}), we initiated the study of the {\em price of fairness} which aims to quantify the impact that the requirement for fairness has on the efficiency of allocations with respect to the social welfare. Formally, it is defined as the worst-case ratio between the maximum social welfare that can be achieved by any allocation and the maximum social welfare over allocations that satisfy particular fairness criteria. In our work, we focused on envy-freeness, proportionality, and another well-known criterion known as equitability, and showed bounds on the price of fairness for divisible and indivisible goods and chores~\cite{CKKK12}. Our work has received a lot of attention and the notion of price of fairness has been considered in many subsequent papers.

When the items are divisible resources, a setting known as {\em cake-cutting}, it is known from previous work that there always exist envy-free and proportional allocations for any number of agents. Such allocations are quite easy to compute when the number of agents is rather small and become harder when more agents come into play. Recently, we took an experimental approach aiming to understand how well-known cake-cutting algorithms for computing fair allocations perform in practice~\cite{KOS22}. We ran many different experiments, where human participants were asked to engage with each other or against automata, and investigated possible manipulations of the algorithms, a question that had only been considered in a small number of previous related papers. Our results indicate that the extent of envy that can arise from strategic behavior is lower in (theoretically) envy-free procedures than proportional and not envy-free ones.

Another problem that recently rose to prominence in fair division of divisible resources is \emph{consensus halving}. In this problem we ask to divide a one-dimensional resource among multiple agents by cutting it into many pieces, and then creating two parts of the pieces such that each agent believes the two parts are of equal value. In \cite{DFHM22} we showed that the problem is \ppa/-complete even for computing a relaxed solution where the agents believe that the difference in value of the two parts is at most a large constant value. Our work has important complexity implications for other major problems in fair division of divisible and indivisible resources, namely discrete ham-sandwich, necklace splitting \cite{FRG22-NS-CH-ham}, pizza sharing \cite{DFM22-PS}, and more. Prior to this, in \cite{DFMS21} we studied the problem of finding an exact solution of consensus halving and showed that its complexity differs tremendously from its approximate version. In particular, the problem becomes \fixp/-hard and belongs \bu/, classes that are not even known to be inside \fnp/.

In contrast to the divisible case, in settings with indivisible items, it is not hard to see that there exist simple instances that do not admit envy-free or proportional allocations (e.g., when there are two agents and one good), and in fact, it is generally hard to even decide when such allocations exist, with the exception of a few special cases such as when each agent is to be assigned only a single item~\cite{gan2019house}. To circumvent such impossibilities, various relaxations of envy-freeness and proportionality have been proposed, including {\em envy-freeness up to one item} (EF1), which demands that every agent does not envy any agent after the hypothetical removal of any item (from the ones given to them), or the stronger {\em envy-freeness up to any item} (EFX), which demands that the envy is eliminated by the removal of any item. EF1 allocations always exist and are easy to compute, but the existence of EFX still remains a mystery and has been shown only for a few special cases, such as when there are at most two possible values for the goods~\cite{amanatidis2021efx}. In our recent survey~\cite{fair-survey}, we discussed important results that have been obtained for these relaxations and other important ones in the discrete fair division literature during the last 10 years. 

We have also studied other interesting settings with different assumptions on the relationships between agents or on the nature of the possible allocations. For example, we have studied the existence and computation of EF1 and EFX allocations for settings where the agents are partitioned into groups and the agents of a group can extract value from the items that are allocated to their group~\cite{kyropoulou2020groups}. We have also considered the case of random allocations and, in particular, the notion of {\em interim envy-freeness}, which aims to provide a tradeoff between the very restrictive notion of (deterministic) envy-freeness and the, perhaps too weak, notion of ex-ante envy-freeness~\cite{CKK21}. 

\paragraph{Future directions.} 
As mentioned above, there are only a few cases for which EFX allocations are known to exist and designing an algorithm that always computes an EFX allocation (even in exponential time) has turned out to be a highly non-trivial task. Experiments on randomly generated data indicate that at least one EFX allocation should always exist, but these do not exclude the possibility of instances that do not admit EFX allocations and have such a structure that coming up with them at random or by hand is quite unlikely. As in utilitarian voting, one possible idea to bypass this bottleneck might be to train a machine learning model to try to identify EFX allocations. Such a model has the potential to lead to new insights, for example, it may identify particular properties shared by all EFX allocations that could be used in the design of an algorithm, or lead to an impossibility result. We plan to investigate this direction in the near future to tackle this enigmatic question in fair division. 

\subsection{Approximate Mechanism Design} \label{sec:md}
As discussed in Section~\ref{sec:games}, in many MAS the participating agents usually have incentives to act selfishly aiming to optimize their personal payoff, thus inducing strategic games. However, in many applications, it is important that we manage to control the behavior of the agents as this can give us a better estimation about the expected system usage and efficiency. This can be done by anticipating possible manipulations by the agents, and thus appropriately specifying the rules of the games they will end up playing. In particular, in {\em mechanism design} the goal is to define the rules such that it is a (weakly) dominant strategy for each agent to act truthfully, that is, no agent has incentive to deviate and report a false input. In some situations this can be coupled together with full efficiency, but usually the truthfulness requirement leads to some loss of societal efficiency, and thus our goal is to design truthful mechanisms that are approximately optimal. In some environments this can be done by using money transfers ({\em mechanism design with money}), whereas in others the use of money is prohibited ({\em mechanism design without money}). 

In settings where monetary transfers are allowed and there are no budget restrictions, it is always possible to simultaneously achieve truthfulness and full efficiency by deploying the well-known Vickrey-Clarke-Groves (VCG) mechanism. Implementing VCG can be hard in certain applications as it demands to identify a welfare-maximizing outcome and then pair it together with particular payments. In some applications, however, it turns out that VCG can be implemented efficiently. One such example is the problem of scheduling tasks to selfish machines in order to minimize the total execution time. In this context, VCG has a very simple interpretation: greedily and myopically allocate each task to a machine that minimizes its processing time (note that this can lead to a sub-optimal completion time). This approach, however, is clearly suboptimal under the more common objective of minimizing the completion time of the schedule, that we have studied in \cite{GK17}. We compute the approximation ratio of VCG when the execution times required for each task are independent random variables, identical across machines. Another simple example is that of single-item auctions, where the VCG mechanism reduces to the simple second-price auction that is truthful and maximizes social welfare (the agent with the highest value for the item obtains it). When the goal is to optimize the revenue in single item auctions rather than welfare, in \cite{CKK16} we showed that no deterministic truthful mechanism can approximate the optimal revenue within a factor better than $63/64 = 98.5\%$  when the participating bidders have correlated valuations; relatively simple optimal mechanisms exist for the case of independent valuations.

In most related literature in mechanism design (and in algorithmic game theory more generally) it is implicitly assumed that the participating agents are perfectly rational in the sense that they always seek to optimize their personal payoffs. But there are also settings with imperfectly rational agents for whom it is not immediate what type of incentives we should provide. {\em Obviously strategyproof} (OSP) mechanisms have been proposed for such cases to require that at each point during the execution where an agent is queried to communicate information, it should be ``obvious'' for the agent what strategy to adopt in order to maximize her utility: roughly speaking, the worst possible outcome after selecting her true type is at least as good as the best possible outcome after misreporting her type. In \cite{DKV20} we considered single-minded combinatorial auctions where each bidder is interested in receiving a specific (known or unknown) bundle of items or a superset of it. We designed novel OSP mechanisms and provided OSP implementations of known truthful mechanisms that are approximately optimal in terms of social welfare. 

There are also many settings where payments are not allowed, but we still want to design truthful mechanisms. The most prominent such setting is that of {\em facility location}. In the simplest model, there is a set of agents with private positions on the line of real numbers and our goal is to decide where to locate a facility such that the total or the maximum distance of the agents from the facility is minimized, while at the same time truthfully eliciting the agent positions. When there are two facilities or more, the agents might also have different {\em preferences} over them leading to the so-called {\em heterogeneous} facility location models. In our work, we have studied a constrained version of the problem with agents that either like or dislike the facilities, and the facilities must be located only within a given feasible region of the Euclidean plane~\cite{kyropoulou2019constrainedFL}. We showed a characterization of the feasible regions for which the optimal solution can be truthfully implemented when the agents can only lie about their preferences or about their locations. In another work, we focused on models with agents that have approval preferences over the facilities (that is, an agent either likes a facility or is indifferent about it) and there are different types of constraints, such as that the agents occupy the nodes of a discrete line graph and the facilities cannot be located at the same node~\cite{kanellopoulos2021discreteFL}, or that we have a limited budget which allows just a subset of the facilities to be built in which case we have to decide which of them to locate and where~\cite{deligkas2022limited}.

In the context of {\em machine scheduling} without payments, one way to incentivize the agents to act truthfully is that of {\em monitoring}. We have studied a version of the problem where the machines can be strategic but are bound by their declarations in the sense that if a machine reports that it needs a particular amount of time to process some jobs assigned to it, then it has to spend at least this much processing time~\cite{KV19}. We showed that by using a novel and particularly harsh notion of monitoring, we can design OSP mechanisms that achieve the same approximation guarantee as the best possible truthful mechanism, and that such a monitoring framework is essential. In the context of {\em fair division}, we have also considered truthful allocations for the fundamental case of two agents and two items \cite{CKKK09}. By characterizing the possible truthful mechanisms, we were able to show an impossibility: deterministic truthful allocations do not minimize the envy between agents. 

Finally, we have also designed strategyproof mechanisms for {\em hybrid social choice} settings, where the objective is to choose a welfare-maximizing outcome when there are both agents with non-monetary preferences (corresponding to voters) over the possible outcomes as well as agents with monetary preferences (corresponding to bidders) \cite{caragiannis2018expert}. Since this problem is a combination of mechanism design with and without monetary transfers, classical solutions like VCG cannot be applied. We focused on the simple but fundamental scenario of one voter and two agents, and provided tight approximation guarantees of the optimal social welfare. We distinguished between mechanisms that use ordinal and cardinal information, as well as between mechanisms that base their decisions on one of the two sides (either the voter or the agents) or both. 

\paragraph{Future directions.} 
The field of approximate mechanism design is broad and contains a number of interesting unexplored directions. For example, a number of questions are left open by our works: how well can we approximate the optimal social welfare in a single-minded combinatorial auction with an OSP mechanism? In what other contexts where payments are not allowed can paradigms like monitoring help incentivize truthful behavior? Beyond the problems we have already considered, it would also be interesting to explore the design of strategyproof machine learning algorithms for settings where the samples might be provided by strategic agents.

\subsection{Multi-Agent Systems in Finance}
The study of financial applications through the lens of algorithmic game theory is a relatively novel research direction that has attracted our interest. \emph{Machine learning} research on finance is more extensive, as it is commonly used in trading strategies and for designing markets, among others. 

\paragraph{Financial network games.}
In \cite{KKZ21} we studied financial networks of firms that can strategize over their payments. In such networks, nodes correspond to firms, and directed labeled edges correspond to debt contracts between them. The existence of cycles in the network indicates that a payment of a firm to one of its lenders might result to some incoming payment. So, if a firm cannot fully repay its debt, then the exact (partial) payments it makes to each of its creditors can affect the cash inflow back to itself. Note that firms are interested in their financial well-being (utility) which is aligned with the amount of incoming payments they receive from the network. We considered a natural set of payment strategies called priority-proportional payments, and we investigated the existence and (in)efficiency
of equilibrium strategies. 

We have also studied a setting where a financial authority offers bailout money to some firm(s) or forgives the debts of others in order to maximize liquidity, and examine efficient ways to achieve this goal, together with game-theoretic questions on the behavior of the firms \cite{KKZ22,KKZ22b}. We investigated the approximation ratio provided by the greedy bailout policy compared to the optimal one, and we studied the computational hardness of finding the optimal debt removal and budget-constrained optimal bailout policy, respectively. From a game-theoretic standpoint, we observed that the removal of some incoming debt might be in the best interest of a firm, if that helps one of its borrowers remain solvent and avoid costs related to default. Therefore, firms aim to maximize their utility by strategically giving up some incoming payments; we studied the existence and quality of pure Nash equilibria, as well as the computational complexity of finding them.

\paragraph{Exploiting machine learning.}
MAS allows the simulation of markets which consist of heterogeneous agents, with differing risk attitudes and differing expectations to future outcomes, in contrast to traditional assumptions of homogeneity and rationality. MAS attempts to explain market behavior, replicate documented features of real-world markets, and allows us to gain insight into the likely outcomes of different regulatory policies. The essence of MAS in financial markets lies in the notion of autonomous agents whose behavior evolves endogenously leading to complex, emergent, system dynamics, which are not predictable from the properties of the individual agents.

In developing a multi-agent system, a key question is: \emph{How do agents learn and adapt their strategies over time?} These ``learning mechanisms'' can be implemented in many ways including the use of different types of machine learning algorithms, such as neural networks or evolutionary algorithms, where each agent's behavior or strategy adapts over time in response to environmental feedback, as we have discussed in \cite{brabazon2020applications}. 

Machine learning has been very popular in financial applications; we have applied it in areas such as trading \cite{adegboye2021machine, kampouridis2017evolving} and derivatives \cite{cramer2018decomposition, cramer2019stochastic}. Past work that has particularly focused on MAS includes studying the \textit{microstructure dynamics} of financial markets. In \cite{kampouridis2012market, kampouridisIRFA2012}, we investigated the market fraction hypothesis (MFH), which states that the fraction of the different types of trading strategies that exist in a financial market changes (swings) over time. Our framework used a genetic programming algorithm to evolve the multi-agent trading strategies and a time-variant self-organizing map to cluster the different types of agents. Our findings led us to reject the hypothesis, as we found that dominant types of trading strategies can remain popular over long time periods with very little swinging present.

Furthermore, in \cite{chen2010microstructure, kampouridis2012microstructure}, we studied the plausibility of the dinosaur hypothesis (DH), which states that because the market's behavior constantly changes, traders need to continuously co-evolve their strategies with the market in order to avoid becoming obsolete or dinosaurs. We again used a multi-agent system that combined evolutionary algorithms and neural networks and provided empirical evidence confirming the DH, thus stressing the need of agents continuously adapting their strategies to the market environment.

Another area that we have worked on is applying machine learning to automated negotiation problems, where we used a genetic programming algorithm to co-evolve negotiation strategies of agents that have different preference criteria, namely price and negotiation speed \cite{kampouridis2013gp}. In cloud/grid computing environments, any delay in acquiring resources is considered an overhead, hence the negotiation agents need to adopt strategies that enable them to not only optimize resource price, but also to reach early agreements. Our work was the first to apply genetic programming to this type of problem and outperformed previous results in the area. 

\paragraph{Future directions.} 
Applying algorithmic game theory techniques to financial networks only recently received attention from the community, hence the field remains to a big extent unexplored. Existing models can be extended by considering more financial instruments (e.g. possibility of claim transfers) and made even more realistic. Computational results could provide insights, e.g. on the inefficiency of equilibria in a random network. Machine learning techniques are very broad and can be applied in this and many other contexts; we propose its use in order to overcome intractability or incomplete information issues also in the previous sections.

\section{Conclusions} \label{sec:conclusion}
Our group combines research expertise that ranges from theoretical computer science to machine learning. Our diverse expertise puts us in the advantageous position of being able to tackle problems in a broad domain of Algorithmic Game Theory, Computational Economics and Finance, both from a theoretical and an applied perspective. This is evident from the extensive (and diverse) list of research problems our group has contributed to, as well as from the numerous joint publications among the group members. We aim to continue advancing research on multi-agent systems for computational economics and finance, and explore the vast landscape in this fascinating area.

\bibliographystyle{alphaurl}           
\bibliography{bibliography}       

\end{document}